\documentclass[a4paper,11pt]{article}

\usepackage{amsmath,amsfonts}
\usepackage{pos}

\newcommand{\ep}{\epsilon}
\newcommand{\Omg}{\Omega}

\title{Phase-space integrals through Mellin-Barnes representation\footnote{Based on Refs.~\cite{Ahmed:2024pxr,Ahmed:2025yrx} by the same authors.}}

\ShortTitle{Phase-space integrals through Mellin-Barnes}

\author[]{Taushif Ahmed\footnote{Speaker}}
\author[]{Syed Mehedi Hasan}
\author[]{Andreas Rapakoulias}

\affiliation[]{Institute for Theoretical Physics, University of Regensburg,\\
93040 Regensburg, Germany}

\emailAdd{taushif.ahmed@ur.de}
\emailAdd{Syed-Mehedi.Hasan@ur.de}
\emailAdd{andreas.rapakoulias@ur.de}

\abstract{%
We compute angular phase-space integrals with three and four 
denominators analytically, working within dimensional regularisation via the
Mellin-Barnes (MB) representation.  The approach converts multifold MB integrals
into real parametric integrals and expresses all results in terms of Goncharov
polylogarithms (GPLs).  For three denominators, all-massless results are obtained
to $\mathcal{O}(\epsilon^2)$ and the single-massive case to $\mathcal{O}(\epsilon)$;
for four denominators, both the massless and single-massive cases are solved to
$\mathcal{O}(\epsilon^0)$.  Integrals with multiple massive momenta follow from a
partial fraction decomposition reducing them to the single-massive case.
Recursion relations relating integrals with higher denominator powers to master
integrals are derived. These are essential ingredients to solving full phase-space integrals. 
}

\FullConference{17th International Symposium on Radiative Corrections:
Applications of Quantum Field Theory to Phenomenology (RADCOR2025)\\
5--10 October 2025\\
Puri, India}

\begin{document}
\maketitle

% ============================================================
\section{Introduction}
% ============================================================

Perturbative QCD predictions at higher order require the evaluation of
real-emission contributions alongside multi-loop virtual corrections.  
At each order, real-emission matrix elements must be integrated over the
corresponding phase space, and this integration generically produces both
infrared singularities---regulated in dimensional regularisation by poles in
$\ep = (4-d)/2$---and finite pieces that enter physical cross sections.  
While the technology for computing multi-loop virtual integrals has matured
substantially over the last two decades, the analytic treatment of
real-emission phase-space (PS) integrals has received comparatively less
attention despite being an equally indispensable ingredient.

In a suitably chosen reference frame, PS integrals in $d$ dimensions
factorize into a process-dependent radial component and a universal angular
part~\cite{Somogyi:2011ir,Anastasiou:2013srw}.  The angular component takes
a particularly clean form: it depends only on the external reference momenta
$\{p_i^\mu\}$ and carries the full infrared pole structure.  Specifically,
with $n$ denominators,
\begin{align}
\label{eq:ang_int}
\Omg_{j_1,\ldots,j_n}(\{p_i^\mu\},d)
  \equiv \int \mathrm{d}\Omega_{d-1}(q)\,
  \frac{1}{(p_1\cdot q)^{j_1}\cdots(p_n\cdot q)^{j_n}}\,,
\end{align}
where $\mathrm{d}\Omega_{d-1}(q)$ is the rotationally invariant measure for a
massless vector $q^\mu$. The on-shell conditions, $p_i^2=0$ or $p_i^2\neq 0$, for the fixed reference momenta
are dictated by the
kinematics of the process at hand.  The integers $\{j_i\}$ depend on the
scattering process and perturbative order; for instance, double real emission
in next-to-next-to-leading order (NNLO) semi-inclusive deep-inelastic scattering (SIDIS) requires at most
$n=2$~\cite{Ahmed:2024owh}, while higher-multiplicity or higher-order
configurations push the requirement toward larger $n$.

The analytic structure of~(\ref{eq:ang_int}) is well understood only for
small $n$.  For $n=1$, the massless case trivially yields the total angular
volume $2\pi^{1-\ep}/\Gamma(1-\ep)$, while the massive case gives a Gauss
hypergeometric ${}_2F_1$~\cite{vanNeerven:1985xr}.  For $n=2$, the massless
result again involves ${}_2F_1$, and one or two massive legs bring in Appell
$F_1$ and Lauricella functions respectively~\cite{vanNeerven:1985xr,Lyubovitskij:2021ges}.
For $n\geq 3$, an exact all-orders-in-$\ep$ representation in terms of the
multivariable $H$-function exists, but expanding it as a Laurent series in
$\ep$ for general kinematics is a genuinely non-trivial problem; neither
a simple hypergeometric closed form nor a straightforward residue sum
suffices in practice.

For $n=3$, the first explicit $\ep$-expansions appeared simultaneously in
refs.~\cite{Ahmed:2024pxr} and~\cite{Haug:2024yfi}: the former obtained
results to $\mathcal{O}(\ep^2)$ and the latter to $\mathcal{O}(\ep)$, but
in different function bases.  Ref.~\cite{Haug:2024yfi} uses Clausen functions,
while ref.~\cite{Ahmed:2024pxr} expresses all results in terms of Goncharov
polylogarithms (GPLs)~\cite{Vollinga:2004sn}.  This distinction has practical
consequences: GPLs admit a fibration basis with well-controlled iterated
integration, so the convolution of the angular part with the radial component
to form the full PS integral is, in principle, tractable analytically.  Clausen
functions do not enjoy this property in the same way.  The small-mass limit of
$n=3,4$ integrals was studied in~\cite{Smirnov:2024pbj} via expansion by
regions, providing an important cross-check on the structure of infrared poles
in the general-mass case.

In refs.~\cite{Ahmed:2024pxr} and~\cite{Ahmed:2025yrx}, we computed $n=3$ and
$n=4$ angular integrals in GPL form using a Mellin-Barnes (MB)-based approach that is both
algorithmic and scalable.  In the $n=4$ case this required evaluating six- and
seven-fold MB integrals analytically in terms of GPLs---to our knowledge the
first results of this type in the literature.  The present contribution
summarises both works.

% ============================================================
\section{Mellin-Barnes representation and expansion strategy}
% ============================================================

The angular integral~(\ref{eq:ang_int}) admits a Mellin-Barnes representation
that encodes the dependence on the kinematic invariants $v_{kl}$ through a
product of $\Gamma$-functions raised to powers $z_{kl}$:
\begin{align}
\label{eq:MB}
\Omg_{j_1,\ldots,j_n}(\{v_{kl}\},\ep)
 &= \frac{2^{2-j-2\ep}\,\pi^{1-\ep}}
         {\prod_{k=1}^n \Gamma(j_k)\,\Gamma(2-j-2\ep)}
   \int_{-i\infty}^{+i\infty}
   \!\!\Bigg[\prod_{k=1}^n\prod_{l=k}^n \frac{\mathrm{d}z_{kl}}{2\pi i}\,
   \Gamma(-z_{kl})(v_{kl})^{z_{kl}}\Bigg]\nonumber\\
 &\quad\times \prod_{k=1}^n\Gamma(j_k+z_k)\;\Gamma(1-j-\ep-z)\,,
\end{align}
where the scalar products among the reference momenta are encoded as
$v_{kl}\equiv (p_k\cdot p_l)/2$ for $k\neq l$ and $v_{kk}\equiv p_k^2/4$,
and $z=\sum_{k,l\geq k}z_{kl}$, $z_k=\sum_{l\leq k}z_{lk}+\sum_{l\geq k}z_{kl}$.
For a massless leg one has $v_{ii}=0$, and the corresponding $z_{ii}$
integration is simply absent.  The number of independent integration variables is equal to the number of distinct non-zero $v_{kl}$. 
For $n=3$ in the all-massless case this number is three ($z_{12}, z_{13}, z_{23}$), 
while it becomes four in the single-massive case. 
For $n=4$, the number increases to six in the all-massless configuration and to seven in the single-massive case.

Because poles of the $\Gamma$-functions can cross the integration contours as
$\ep\to 0$, a direct Taylor expansion of the integrand is generically
ill-defined.  We proceed in four steps:

\begin{enumerate}
\item \textbf{Analytic continuation - }  following~\cite{Smirnov:1999gc,Tausk:1999vh,Czakon:2005rk},
  we track the motion of $\Gamma$-function poles as $\ep$ varies toward zero.
  Each time a pole of type $\Gamma(a+z_{kl})$ or $\Gamma(a-z_{kl})$ crosses the
  contour for $z_{kl}$, we collect the corresponding residue---a lower-fold MB
  integral---and update the remaining integral.  Residue terms may themselves
  contain contour-crossing singularities and are treated iteratively until the
  integrand is safe to expand.

\item \textbf{$\ep$-expansion - }  once the contours are clear of crossings, the
  integrand is expanded in $\ep$.  Each coefficient in the Laurent series is
  a (generally simpler) MB integral, typically with fewer integration variables
  or with gamma-function arguments whose real parts are manifestly positive.

\item \textbf{Reduction to real parametric integrals - }  after analytic
  continuation and expansion, the resulting MB integrals are
  \emph{balanced}: the number of $\Gamma(a+z_j)$ factors equals the number of
  $\Gamma(a-z_j)$ factors for each variable $z_j$ individually.  This allows
  the $\Gamma$-products to be assembled into Euler beta functions, which are
  then replaced by their integral representations.  The remaining MB
  integrations over the $\tilde z_l$ are carried out via
  $\int_{-i\infty}^{+i\infty}(dz/2\pi i)\,A^z = \delta(1-A)$, collapsing each
  one to a $\delta$-function constraint and leaving ordinary integrals over
  parameters $x_i\in[0,1]$ or $[0,\infty)$.

\item \textbf{Express in terms of GPLs - }  the resulting integrands are rational functions
  and logarithms of polynomials in the $x_i$, naturally suited for iterated
  integration into GPLs~\cite{Vollinga:2004sn}.  Whenever an integration variable
  appears non-linearly inside a GPL weight, we rewrite that GPL via its integral
  representation in terms of lower-weight ones and shift the variable to the
  rightmost (outermost integration) position using in-house algorithms, until
  the variable sits exclusively in the rightmost argument of every
  GPL~\cite{Duhr:2019tlz}.  Quadratic radicands $\sqrt{ax^2+bx+c}$ obstructing
  integration are rationalised by $x\to(b+2c\eta)/(c\eta^2-a)$; the
  transformation converts all arguments and the Jacobian to rational functions
  of $\eta$ without introducing new square-root obstacles.  All final
  expressions are verified numerically.
\end{enumerate}

% ============================================================
\section{\texorpdfstring{Three-denominator results~\cite{Ahmed:2024pxr}}{Three-denominator results}}
% ============================================================

\subsection*{Massless case}

For $n=3$ with all massless $p_i$, the integral depends on three kinematic
invariants $v_{12},v_{13},v_{23}$, and the starting point is the threefold MB
representation~(\ref{eq:MB}).  We absorb a conventional normalisation factor
into
$I^{(0)}_{j_1,j_2,j_3}=C_\ep\,\Omg_{j_1,j_2,j_3}$
with
$C_\ep = 2^{-1+2\ep}\pi^\ep\Gamma(1-2\ep)/\Gamma(1-\ep)$;
the superscript $(0)$ labels the all-massless configuration.

Setting $j_i=1$---the denominator powers that arise in NNLO and N$^3$LO
double- or triple-real corrections---the analytic continuation of the threefold
MB integral identifies pole crossings and generates a hierarchy of lower-fold
residue integrals.  Collecting all contributions at each order in $\ep$, we
find four distinct real-parameter integral types at $\mathcal{O}(\ep)$ and
eight at $\mathcal{O}(\ep^2)$, of which four are genuinely new at the higher
order.  Crucially, no square roots appear over the integration variables at
either order; the radicand rationalisation step is not needed here.  Carrying
out the iterated integrations following step~4 above, $I^{(0)}_{1,1,1,1}$ is
fully expressed in terms of GPLs up to $\mathcal{O}(\ep^2)$---the first such
result in the literature.  The explicit expression is provided in the ancillary
file \texttt{3prop0mass.m}~\cite{Ahmed:2024pxr}.

The pole structure is consistent with the expected collinear singularity: a
single $1/\ep$ pole, with no double pole.  This follows from the fact that the
$n=3$ massless angular integral can develop at most a single collinear
divergence from the $q\parallel p_i$ configuration.  Results to
$\mathcal{O}(\ep^2)$ are necessary in higher-order computations where the
angular integral appears at a lower-loop order and must be expanded two powers
deeper in $\ep$ to produce finite contributions after $\ep$-pole
cancellations~\cite{Ahmed:2024owh}.

\subsection*{Single-massive case}

When $p_1^\mu$ carries mass ($p_1^2\neq 0$, so $v_{11}\neq 0$), an additional
integration variable $z_{11}$ enters and the MB representation~(\ref{eq:MB})
becomes fourfold.  After analytic continuation and expansion, eleven distinct
MB integrals appear at $\mathcal{O}(\ep)$.  Ten of these are handled by the
same procedure as in the massless case, with no square roots over integration
variables.  The eleventh one contains a quadratic radicand.  Applying the substitution
$x\to(b+2c\eta)/(c\eta^2-a)$ rationalises the square root, after which the
remaining integrals are of standard GPL type.  The result
$I^{(1)}_{1,1,1,1}$---superscript $(1)$ denoting one massive leg---is
expressed in terms of GPLs to $\mathcal{O}(\ep)$ and provided in
\texttt{3prop1mass.m}~\cite{Ahmed:2024pxr}.

\subsection*{Multi-massive cases and partial fractions}

When two or three of the $p_i^\mu$ are massive, a direct MB treatment would
increase the number of integration variables substantially.  However, the
angular integral is linear in parameter space when the reference momenta are
expressed in a suitable coordinate system, which permits a partial fraction
(PF) decomposition of the propagators.  Exploiting this linearity, every
multi-massive $n=3$ integral can be expressed as a sum of single-massive
ones~\cite{Ahmed:2024pxr}.  For the double-massive case the PF coefficients
involve the combination
$\lambda_\pm = [2v_{12}-4v_{11} \pm \sqrt{4v_{12}^2-16v_{11}v_{22}}]/(4v_{12}-4v_{11}-4v_{22})$,
which originates from the two-point splitting lemma of
ref.~\cite{Lyubovitskij:2021ges}.  The triple-massive case follows by iterated
application.  Explicit formulae are given in ref.~\cite{Ahmed:2024pxr}.

\subsection*{Recursion relations}

Angular integrals with higher denominator powers ($j_i>1$) are not
independent: differentiating with respect to kinematic invariants generates
linear relations among integrals with shifted indices,
\begin{align}
  \frac{\partial}{\partial v_{kl}}I^{(n)}_{j_1,j_2,j_3}
  = \sum_{i_1,i_2,i_3} C^{(i_1 i_2 i_3)}_{kl}\,I^{(n)}_{i_1,i_2,i_3}\,,
\end{align}
entirely analogous to integration-by-parts (IBP)
identities~\cite{Chetyrkin:1981qh,Laporta:2000dsw} for loop integrals.  By
iterating these six linear equations, any $I^{(n)}_{j_1,j_2,j_3}$ with
arbitrary $j_i$ reduces to the set $\{I^{(n)}_{1,1,1}\}$ as master integrals.
The complete set of six relations is provided in \texttt{recursion.m}~\cite{Ahmed:2024pxr}.

\subsection*{Combining with the radial part}

The full PS integral is obtained by convoluting the angular result with the
process-dependent radial component over a single parametric integration, where
the integration variable is a dimensionless ratio of Mandelstam invariants.
For the massive angular integrals, the $\ep$-expansion coefficients are regular
along the integration path, so the expanded form can be used order by order
without difficulty.

For the massless angular integrals, soft singularities also appear within the
individual $\ep$-coefficients of the angular part, obstructing a naive
term-by-term treatment.  Since these singularities are soft in origin, they
exponentiate and factorise from the full angular result to all orders in $\ep$,
in direct analogy with soft divergence exponentiation in
QCD~\cite{Ahmed:2024pxr}.  We have verified this factorisation explicitly for
the $n=2$ case~\cite{Ahmed:2024owh}.  Correctly separating the soft factor via
pole subtraction renders the remaining angular piece safe to expand order by
order, reducing the final step to a one-dimensional parametric integral over an
analytic expression.

% ============================================================
\section{\texorpdfstring{Four-denominator results~\cite{Ahmed:2025yrx}}{Four-denominator results}}
% ============================================================

\subsection*{Massless case}

For $n=4$ with all massless $p_i$, there are six kinematic invariants
$v_{12},v_{13},v_{14},v_{23},v_{24},v_{34}$, and the MB representation is a
sixfold integral over $\{z_{12},z_{13},z_{14},z_{23},z_{24},z_{34}\}$.  After
the four-step procedure of section~2, the analytic continuation and
$\ep$-expansion produce ten real-parameter integrals $I_1,\ldots,I_{10}$ (with
integration ranges $v_1\in[0,v_{12}]$ and $x_i\in[0,1]$).  Evaluating these
via iterated integration yields $I^{(0)}_{1,1,1,1} = C_\ep\,\Omg_{1,1,1,1}$
up to $\mathcal{O}(\ep^0)$.

The $\ep^{-1}$ pole is fully determined by the collinear singularity structure
and takes the manifestly $S_4$-symmetric form
\begin{align}
\label{eq:pole0}
  I^{(0),-1}_{1,1,1,1}
  = -\frac{\pi}{8}\cdot\frac{1}{4!}
    \sum_{\sigma\in S_4}
    \frac{v_{\sigma(1)\sigma(2)}\,v_{\sigma(1)\sigma(3)}\,v_{\sigma(2)\sigma(3)}}
         {\displaystyle\prod_{1\leq i<j\leq 4}v_{ij}}\,,
\end{align}
in agreement with the small-mass limit of ref.~\cite{Smirnov:2024pbj}.  The
symmetrisation over permutations of the four massless legs is exactly what one
expects from the underlying rotational invariance of the angular measure.  

The finite part $I^{(0),0}_{1,1,1,1}$ involves weight-2 GPLs whose alphabet
contains 17 letters.  Nine of these are rational in the kinematic invariants,
while eight are square-root-valued, coming in pairs
\begin{align}
  l_{8,13} = \frac{X_1 \mp \sqrt{Y_1}}{2v_{14}v_{23}}\,,\quad
  Y_1 = v_{14}^2v_{23}^2 - 2v_{14}v_{23}(v_{13}v_{24}+v_{12}v_{34}) + X_6^2\,,
\end{align}
where $X_6 = v_{13}v_{24} - v_{12}v_{34}$.  These square roots do \emph{not}
arise during the real-parameter integrations themselves; they emerge only when
the iterated integrals are recast in the GPL basis, where they play the role of
branch points of the weight function.  They are the $n=4$ analogue of the Gram
determinant square roots familiar from two-loop Feynman integral calculations.
The full analytic expression is in \texttt{massless\_ep0.m}~\cite{Ahmed:2025yrx}; numerical
verification at three independent phase-space points against direct MB
evaluation confirms our result~\cite{Ahmed:2025yrx}, with a $\sim 1800\times$
speedup: evaluation via \textsc{GiNaC} takes ${\sim}1$\,s versus ${\sim}30$\,min
for direct numerical MB integration.

\subsection*{Single-massive case}

Adding one massive leg $p_1^\mu$ introduces $z_{11}$, making the MB
representation sevenfold.  The analytic continuation and expansion now produce
fifteen real-parameter integrals.  The $\ep^{-1}$ pole,
\begin{align}
\label{eq:pole1}
  I^{(1),-1}_{1,1,1,1}
  = -\frac{\pi}{8}\cdot\frac{1}{2}
    \sum_{\sigma\in S_3}
    \frac{v_{1\sigma(2)}\,v_{1\sigma(3)}\,v_{\sigma(2)\sigma(3)}}
         {\displaystyle\prod_{1\leq i<j\leq 4}v_{ij}}\,,
\end{align}
is now symmetrised only over permutations of the three massless legs
$\{2,3,4\}$, reflecting the broken $S_4$ symmetry once $p_1^\mu$ is singled
out as massive.  This result is consistent with refs.~\cite{Smirnov:2024pbj,Salvatori:2024nva}.
The finite part involves weight-2 GPLs with an alphabet of 11 letters (again
including square-root letters), provided in \texttt{one\_mass\_ep0.m}~\cite{Ahmed:2025yrx}.

The successful evaluation of the six- and sevenfold MB integrals in terms of
GPLs constitutes, to our knowledge, the first such analytic result in the
literature.  The method is fully algorithmic and encounters no new conceptual
obstacles as the number of folds increases; the principal challenge is
computational rather than structural.

\subsection*{Double-, triple-, and quartic-massive cases}

As in the $n=3$ case, all multi-massive configurations are handled by
extending the PF decomposition to $n=4$.  For two massive legs $p_1^\mu$ and
$p_2^\mu$ the result is
\begin{align}
\label{eq:double_mass_pf}
  I^{(2)}_{1,1,1,1}
  = \lambda_{12}\,I^{(1)}_{1,1,1,1}(v_{11},\ldots)
  + \bar\lambda_{12}\,I^{(1)}_{1,1,1,1}(v_{22},\ldots)\,,\quad
  \lambda_{ij} = \frac{2v_{ii}-v_{ij}-\sqrt{v_{ij}^2-4v_{ii}v_{jj}}}
                      {2(v_{ii}-v_{ij}+v_{jj})}\,,
\end{align}
where $\bar\lambda_{ij}=1-\lambda_{ij}$, and the new scalar products in the
arguments of the two single-massive integrals on the right are given
explicitly in terms of the original $v_{kl}$.  The same procedure applied
iteratively yields explicit formulae for the triple- and quartic-massive cases,
each expressed as a sum of single-massive integrals with shifted arguments.
Full expressions and the corresponding new scalar products are given
in~\cite{Ahmed:2025yrx}.  Numerical evaluation of all mass configurations
takes well under a second.

\subsection*{Recursion relations}

Differentiating $I^{(n)}_{j_1,j_2,j_3,j_4}$ with respect to $v_{kl}$ generates
recursion relations among four-denominator integrals with shifted denominator
powers, entirely analogous to those in the $n=3$ case.  These hold for all mass
configurations and reduce any integral with arbitrary $j_i$ to the master
integrals computed above.  Details are given in~\cite{Ahmed:2025yrx}.

% ============================================================
\section{Summary and outlook}
% ============================================================

We have presented an MB-based method for computing angular PS integrals with
three and four propagator denominators, expressing all results as Laurent series
in $\ep$ in terms of GPLs.  The pipeline---analytic continuation,
$\ep$-expansion, balanced-MB-to-real-integral conversion, and GPL
reduction---is fully algorithmic and was implemented for $n=3$
in~\cite{Ahmed:2024pxr} and extended to $n=4$ in~\cite{Ahmed:2025yrx}.

For $n=3$, the first GPL results were obtained: all-massless kinematics to
$\mathcal{O}(\ep^2)$ and the single-massive case to $\mathcal{O}(\ep)$, with
all other mass configurations covered by PF decomposition and a complete set
of recursion relations.  The soft singularity structure when combining angular
and radial parts was identified and handled.  For $n=4$, the massless and
single-massive results to $\mathcal{O}(\ep^0)$ required evaluating six- and
seven-fold MB integrals analytically---the first in the literature.  The
GPL representation provides a practical benefit in numerical evaluation: with
\textsc{GiNaC} the computation takes ${\sim}1$\,s versus ${\sim}30$\,min for
direct MB integration.

The GPL representation of the angular results is directly suited for
convolution with process-dependent radial components, since GPLs close under
iterated integration.  This is the decisive advantage over the Clausen-function
representation of ref.~\cite{Haug:2025sre}: the final step of performing the
radial integration can in many cases be carried out analytically, yielding
fully analytic PS integrals.  When analytic evaluation of the radial integral
is not feasible, the one-dimensional structure of the remaining integral still
permits high-precision numerical computation with ease.  Both routes are 
 pursued in the context of NNLO SIDIS~\cite{Ahmed:2024owh}.

When the relevant function space exceeds that spanned by GPLs---a situation
familiar from multi-loop integral calculations at higher perturbative orders or
with multiple mass scales---the same method generates results in terms of
iterated integrals, which retain the key structural properties needed for the
radial convolution.  The extension to $n\geq 5$ follows without new conceptual
obstacles, with the evaluation of higher-fold MB integrals as the only
practical challenge.

\acknowledgments

The work of T.A.\ and A.R.\ is supported by the Deutsche Forschungsgemeinschaft
(DFG) through Research Unit FOR2926, ``Next Generation Perturbative QCD for Hadron
Structure: Preparing for the Electron-Ion Collider'', project number 409651613.
The authors thank the organisers of RADCOR 2025 for the kind invitation and warm
hospitality.

% ============================================================
% Bibliography
% ============================================================
%% NOTE: compile with pdflatex only (do NOT run BibTeX).
%% If using Overleaf, set the compiler to "LaTeX" or disable BibTeX.
%% Any "missing entry" or "references.bib not found" errors are from
%% a stale .aux file; delete it and recompile.

{\sloppy

} % end \sloppy

\end{document}